\documentstyle[12pt,epsf]{article}

\date{April 30, 1996}
\newcommand{\tenrm}{\small}
\newcommand{\be}{\begin{equation}}
\newcommand{\ee}{\end{equation}}
\newcommand{\bea}{\begin{eqnarray}}
\newcommand{\eea}{\end{eqnarray}}

\begin{document}
\title{ Dynamical flavour dependence of static 
heavy meson decay constants on the lattice}

\author{G.\ M.\ de Divitiis, R.\ Frezzotti, M.\ Masetti 
and R.\ Petronzio \\
\small Dipartimento di Fisica, Universit\`a di Roma {\em Tor Vergata} \\
\small and \\
\small INFN, Sezione di Roma II \\
\small Viale della Ricerca Scientifica, 00133 Roma, Italy \\
\medskip
}

\maketitle

\begin{abstract} 
We study the dynamical flavour dependence of the lattice estimates
of the heavy meson decay constant in the static limit. We perform
the analysis by considering negative flavour numbers and
by extrapolating the results to positive values.
We observe a sizeable flavour dependence which increases the 
quenched estimates.
\end{abstract}

\vfill

\begin{flushright}
  {\bf ROM2F-96-16 ~~~~}\\
\end{flushright}

\newpage

The knowledge of leptonic decay constants of heavy mesons enters the determination of the Cabibbo-Kobayashi-Maskawa mixing angles.
In particular, the value of the $B$ meson decay constant constraints
the shape of the unitarity triangle and sets the size of possible CP violation
in $B$ decays.
Non 
perturbative estimates of this quantity
can be obtained from lattice calculations which are still confronted with various 
sources of systematic errors. The
presence in the dynamics of a heavy meson system of two very different
mass scales implies that
reasonable values for exploring the $B$ meson dynamics with enough
resolution for the heavy quark propagation and enough volume for the
heavy meson wave function are a lattice size of the order of $1.5$ fm and a 
lattice spacing of order of $0.02$ fm. These values lead to about $75$ 
lattice points, beyond present computer capabilities.

Estimates for the physical region of the heavy quark mass
are obtained by interpolating between the
results which are obtained in the charm quark mass region and those
deriving from an
expansion of the fermion action in the inverse of the heavy quark mass. 
The first term in the expansion is the so called static limit \cite{eichten_hill} and
corresponds to the approximation where the heavy quark does not propagate
in space.
An important source of possible systematic errors of such computations is the
quenched approximation: the inclusion of dynamical quark loops
could alter the quenched estimates and vanify the efforts
of reducing the statistical fluctuations within the quenched approximation.
In this letter we make an estimate of the
unquenching effects on
the pseudoscalar decay constant in the static limit with
the ``bermion'' method \cite{Noi}, i.e. we extrapolate from negative
to positive flavour numbers
the results obtained at fixed
renormalized parameters.

On the lattice and in the static limit, the correlator of two local 
heavy-light currents is given by:

\be
G(t) = 
\frac{1}{L^3T}
\sum_{x,y} \delta^3_{\vec{x},\vec{y}} \delta_{t_y,t_x+t}
\langle {\mbox {Tr}} \big[ P(x,y)\frac{1-\gamma_0}{2}S_q^\dagger(x,y)\big] 
\rangle 
\label{corr}
\ee
where $S_q$ is the light quark propagator, $P(x,y)$ 
is the Polyakov line from $x$ to $y$, and $L$ and $T$ are the 
space and time sizes respectively. 

At large times there is a single state dominating the correlation and one 
can extract the matrix element of the heavy-light current between such a state and the vacuum:

\be
G(t)\quad \stackrel{t~ large}{\longrightarrow}  \quad 
Z_L^2 e^{-\Delta E t}\label{corrlarget}
\ee

The decay constant in the static limit can be derived from $Z_L$: 
\be
f_P\sqrt{M_P} = \sqrt{2}Z^{Ren}Z_L a^{-3/2}
\ee
where $Z^{Ren}$ is the suitable renormalization constant of the lattice 
current. 

In order to simulate theories with negative flavours we 
introduce a ``bermion'' action \cite{Noi} which is given by 
$S_g[U] + \sum_i S_{\phi_i}[U,\phi_i]$, where $S_g$ is the standard
Wilson action for the gauge sector and the second term is
a sum over flavour indices of

\be
S_{\phi_i}[U,\phi_i] = \sum_{x} |[Q\phi_i](x)|^2
 \label{bermact}
\ee
with $\phi_i(x)$ the bermion field of flavour $i$. 
For the lattice Dirac operator we follow the standard Wilson formulation:
\bea
[Q\phi_i](x) = \frac{1}{2\kappa}\gamma_5 \phi_i(x)
- \frac{1}{2}\gamma_5\sum_{\mu=0}^3
U_\mu(x)(1-\gamma_\mu)\phi_i(x+\mu) \nonumber \\
- \frac{1}{2}\gamma_5\sum_{\mu=0}^3 
U^\dagger_\mu(x-\mu)(1+\gamma_\mu)\phi_i(x-\mu) 
\eea
where $\kappa$ is the Wilson hopping parameter related to the bare mass.

The square of the Dirac operator is needed
for the Montecarlo update of the bermion fields and implies that
the two point function of these fields represents the inverse of such a square.
The usual inverse Dirac operator can be obtained by remultiplying the
bermion field correlations with the Dirac operator. 

The correlation $G(t)$ can be estimated with
bermion field averages, in a way analogous
to the one followed in \cite{noiquenched} for the quenched case.
In the quenched case, the ``bermion'' fields are not dynamical,
 they are
thermalized in a frozen gauge configuration and provide a Monte Carlo
inversion of the Dirac operator.

According to the method presented in \cite{noiquenched}, we define two
correlators, the first corresponding to eq.\ \ref{corr}:

\be
G(t) = 
\frac{1}{n_bL^3T} \sum_{i=1}^{n_b}
\sum_{x,y} \delta^3_{\vec{x},\vec{y}} \delta_{t_y,t_x+t}
\langle  [Q\phi_i]^\dagger(x)\gamma_5 P(x,y)\frac{1+\gamma_0}{2}\phi_i(y)
\rangle 
\label{corrQ}
\ee
and a second one corresponding to propagators which are the inverse of
 the square of the Dirac operator:

\be
G^{(Q^2)}(t) = 
\frac{1}{n_bL^3T} \sum_{i=1}^{n_b}
\sum_{x,y} \delta^3_{\vec{x},\vec{y}} \delta_{t_y,t_x+t}
\langle  \phi_i^\dagger(x)P(x,y)\frac{1+\gamma_0}{2}\phi_i(y)
\rangle 
\label{corrQ2}
\ee
where $n_b$ is the number of bermion flavours over which the sum is performed. 

The statistical fluctuations of these operators are reduced by
the ``one link integral'' method \cite{parisi_pr}: at 
fixed bermion configuration
one replaces the link by its average in the surrounding
bermion and gauge configuration.

As already noticed earlier \cite{MdaggerM} the correlation 
function constructed from 
the inverse of the square of the Dirac operator in eq.\ \ref{corrQ2}
projects more precociously on the lowest
lying state. We have therefore extracted the energy shift $\Delta E$ 
from this correlation and then fixed its value in the fit of the 
canonical correlator of eq.\ \ref{corrQ}.
The simulations were performed on a 25 Gigaflop machine of the APE series.
The update procedure was for the gauge sector a Cabibbo--Marinari 
pseudo--heatbath \cite{Cab} followed by three overrelaxation sweeps 
and for the bermions a heat bath followed by ten 
overrelaxation sweeps \cite{LuscherAlgorithm}.
The measurements were performed every five sweeps:
we have checked that the autocorrelation time, defined as 
$\tau = \sigma_{true}^2/(2\sigma_{na\ddot{\i} f}^2)$, where $\sigma_{true}$ is 
evaluated by collecting measurements into bins and $\sigma_{na\ddot{\i} f}$ 
is evaluated by considering each measurement
as an independent one, is around 0.5  for the  
correlations $G(t)$ and around 1 for $G^{(Q^2)}(t)$.
According to the procedure of extrapolating at 
fixed renormalized quantities, the values of the $\beta$ and $\kappa$
in each simulation with a given number of
bermions were chosen to
match the $\rho$ and $\pi$ masses at the corresponding quenched values. 
We have explored two values of the ratio $R_2 = m_\pi^2/m_\rho^2$, 0.5 and
0.6, with a lattice spacing corresponding to the value
$\beta = 5.7$ of the quenched case and 
appropriate values of $\kappa$. 

We have made simulations  
with $n_f=-2$, $-4$ and $-6$ for $R_2 = 0.5$ and 
$n_f= -2$, $-4$, $-6$ and $-8$ for $R_2 = 0.6$.
The quenched values are those of ref. \cite{noiquenched} 
with a slight different statistics in the case of $\kappa = 0.163$.
The errors are evaluated as follows: we form clusters of 200 measurements. 
On these clusters we apply a jackknife algorithm to estimate
the errors of the correlations. Each ``jackknife cluster'',
defined by single elimination of a cluster from the total sample is fitted
with the MINUIT minimization program. From the spread of the fits 
on each jackknife
cluster we extract the errors that we quote for our results. 
For the operator of eq.\ \ref{corrQ2} we have checked that
the results of one and two mass fits are generally compatible while
for the operator of eq.\ \ref{corrQ} a two mass fit is necessary with the
lowest mass fixed by the fit of the other operator. 
We quote separately the statistical error of the
fit and the error deriving from the uncertainty
by which the correlation $G^{(Q^2)}(t)$ fixes the value of the
lowest energy shift.
All the results have been obtained on a $16^3\times 32$ lattice and the 
correlations have been taken in all directions, with a caveat.
The correlations in the space directions, for values of the separation close
to the boundary (beyond 11-12) may suffer from finite size effects due to the
presence of dynamical ``bermions''. Indeed,
the light quark can connect a point in the original lattice with a point
sitting in the backward copy,
while the heavy quark is explicitely propagated only in the forward
direction and sits always in the original lattice.
The result is a path wrapping around the lattice. In the
quenched approximation and in the confined phase, these paths average to zero,
but they are in general non zero in the unquenched case.
The fermion determinant may create a fermion loop 
wrapping around the lattice which, through its correlation
with the one from the heavy-light propagator can
give to the latter a non zero expectation value. 
The effect is visible beyond the statistical errors for $t > 10$.
We have then included in the fits the points from the correlation 
in the space directions up to $t = 10$ only.
For the time direction which is twice longer the effect is negligible
for the maximum values ($t=14$) of the time distance where the signal is still
appreciable.

Figures 1 and 2 give the flavour dependence of the static
decay constants for the two values of $R_2$. In Tables 1 and 2 we 
present our detailed numerical results for the energy shift, for
$Z_L$ and for the parameters of our simulations including the matched
values of $\pi$ and $\rho$ masses.
The extrapolation of the results to $n_f= +3$
has been attempted with a linear and a quadratic fit
 with the results collected in table 3. 
While the former varies with the number
of negative flavours included in the fit, the latter is more stable
against this variation and leads to values which are compatible within
our errors. This applies also to the case of 
$R_2=0.6$ where we have explored the flavour dependence up to
$n_f = -8$.
In order to take into account the error arising from
an imperfect matching of some simulations, we have extrapolated the 
values of  $Z_L$ corrected by the 
factor $(m_{\rho}^{n_f=0}/m_{\rho}^{n_f})^{3/2}$.
The unquenched estimates at  $R_2=0.5$ and $R_2=0.6$
can be further extrapolated in 
the quark mass (i.e. in the inverse of the Wilson hopping
parameter) to the
chiral limit where a physical values of $f_B^{static}$ can be obtained
by using the experimental value of the $\rho$ mass as an input.

Our final results for the unquenched $f_B^{static}$ at a lattice
spacing corresponding to $\beta = 5.7$ of the quenched case, 
extrapolated to the chiral limit with $n_f=3$ and not corrected for the
renormalization constant is:

$$ f_B^{static}/Z^{Ren} =  Z_L \sqrt{2\over M_B} \: ({{m_{\rho}^{phys} }
\over m_{\rho}^{lattice}})^{3/2} = 0.66(7) {\mbox GeV} $$

to be compared with the quenched result:

$$ f_B^{static}/Z^{Ren} = 0.50(3) {\mbox GeV} $$

The unquenched estimate for
the ratio $f_{B_s}^{static}/f_{B_u}^{static}$ cannot be obtained directly
from the
quark mass dependence of our results. Indeed, in our simulations
we have identified the valence and the sea quark masses. This
is appropriate for the light quark spectrum, while for
the strange quark spectroscopy one should keep the two types of masses
distinct. We defer such a refined study to a future work.

Including dynamical fermion loops increases the static
decay constants, an effect already claimed by the MILC
collaboration which uses dynamical staggered quarks \cite{MILC_staggered}.

The results discussed so far are for the bare quantities, not corrected
for the appropriate renormalization constants. Their inclusion partially 
compensates the observed flavour dependence, 
but the effect can be estimated to be smaller than the 
increase of $Z_L$ due to the unquenching.
Indeed by using the bare couplings in the one loop
expression for $Z^{Ren}$, 
the percentage of the unquenching effects is {\it decreased} by a few units.
By using an improved
perturbation theory, i.e. a coupling constant in a Parisi
\cite{parisi_coupling} scheme, the effect is larger and the
percentage decreases by about ten units.
The extrapolated values of $Z^{Ren}$ in the
Parisi coupling scheme are given
in Table 3.

A sizeable flavour dependence survives the corrections
of the renormalization constant, quantifies into a 10-15 \%
increase in the intermediate quark
mass region that we have explored and becomes about 20\% in the
chiral limit.

\newpage

\begin{table}
\begin{center}
\begin{tabular}{|| c | c | c | c | c | c | c | c ||}\hline 
 $n_f$ & $n_{meas}$ & $\beta$ & $\kappa$ & $\Delta E$ & $Z_L$        & $m_\pi$  & $m_\rho$  \\ 
\hline 
  $-8$ & 6000       & 6.99    & 0.153    & 0.742(2)   & 0.440(2)[7]  & 0.563(5) & 0.712(10)  \\ 
  $-6$ & 5000       & 6.71    & 0.1548   & 0.751(3)   & 0.466(2)[8]  & 0.559(6) & 0.716(9)   \\ 
  $-4$ & 3000       & 6.4     & 0.157    & 0.770(3)   & 0.515(4)[9]  & 0.552(3) & 0.710(6)  \\ 
  $-2$ & 4400       & 6.06    & 0.1598   & 0.789(2)   & 0.573(2)[7]  & 0.566(4) & 0.714(8) \\ 
  ~$0$ & 9200       & 5.7     & 0.163    & 0.811(1)   & 0.648(4)[4]  & 0.562(2) & 0.719(4)  \\ 
\hline 
\end{tabular}
\end{center}
\caption{Results for $\Delta E$ and $Z_L$ at $R_2 = 0.6$ and
 the matched values of $m_\pi$ and $m_\rho$. The second column gives
the number of measurements in each case. For $Z_L$ we separate 
the statistical error () from the one coming from the determination 
of $\Delta E$ []. }
\label{TAB_1}
\end{table}

\begin{table}
\begin{center}
\begin{tabular}{|| c | c | c | c | c | c | c | c ||}\hline 
 $n_f$ & $n_{meas}$ & $\beta$ & $\kappa$ & $\Delta E$ & $Z_L$        & $m_\pi$  & $m_\rho$  \\ 
\hline 
  $-6$ & 3600       & 6.785   & 0.156    & 0.733(2)   & 0.440(2)[5]  & 0.477(8) & 0.67(1)   \\ 
  $-4$ & 3400       & 6.463   & 0.158    & 0.744(2)   & 0.465(7)[6]  & 0.464(4) & 0.658(8)  \\ 
  $-2$ & 5000       & 6.1     & 0.161    & 0.762(3)   & 0.520(10)[9] & 0.467(2) & 0.662(5)  \\ 
  ~$0$ & 6200       & 5.7     & 0.165    & 0.786(2)   & 0.590(7)[7]  & 0.457(3) & 0.659(8)  \\ 
\hline 
\end{tabular}
\end{center}
\caption{The same quantities of table 1 for $R_2 = 0.5$.}
\label{TAB_2}
\end{table}

\begin{table}
\begin{center}
\begin{tabular}{| c | l | l | l | l | l | l | c |}\hline
 \# of  & \multicolumn{7}{c}{{\large $R_2=0.5$}}   \vline\\
\cline{2-8} 
  points     & \multicolumn{2}{c}{$Z_L(n_f=3)$} \vline& \multicolumn{2}{c}{$Z^{Ren}(n_f=3)$ } \vline&
         \multicolumn{2}{c}{$Z_L  Z^{Ren}(n_f=3)$} \vline&
         ${{Z_L  Z^{Ren}({n_f=3})} \over Z_L  Z^{Ren}({n_f=0})}$ \\
\cline{2-8} 
& parabolic & linear & parabolic & linear & parabolic & linear  &  parabolic\\
\hline 
 2 & -           & 0.702(38) & -    & 0.662 & -        & 0.473(29) & -    \\ 
 3 & 0.744(55)   & 0.681(22) & 0.647 & 0.670 & 0.490(42) & 0.465(17) & -   \\ 
 4 & 0.730(23)   & 0.655(17) & 0.649 & 0.678 & 0.485(18) & 0.453(14) & 1.16(4)    \\ 
\hline 
 & \multicolumn{7}{c}{{\large$R_2=0.6$}}   \vline\\
\cline{2-8} 
       & \multicolumn{2}{c}{$Z_L(n_f=3)$} \vline& \multicolumn{2}{c}{$Z^{Ren}(n_f=3)$ } \vline&
         \multicolumn{2}{c}{$Z_L  Z^{Ren}(n_f=3)$} \vline&
         ${{Z_L  Z^{Ren}({n_f=3})} \over Z_L  Z^{Ren}({n_f=0})}$ \\
\cline{2-8} 
& parabolic & linear & parabolic & linear & parabolic & linear  &  parabolic\\
\hline 
 2 & -        & 0.742(25)  & -    & 0.668 & -        & 0.503(20) & -    \\ 
 3 & 0.779(42) & 0.730(17)  & 0.659 & 0.673 & 0.519(33) & 0.499(13) & -    \\ 
 4 & 0.760(19) & 0.730(13)  & 0.659 & 0.678 & 0.506(15) & 0.504(11) & -    \\ 
 5 & 0.778(12) & 0.718(12)  & 0.661 & 0.684 & 0.522(9)  & 0.498(9)  & 1.14(2) \\ 
\hline 
\end{tabular}
\end{center}
\caption{We give the results for the values of $Z_L$ extrapolated to $n_f=3$
at $R_2=0.5$ and $R_2=0.6$ as a function of the number of points 
used in the extrapolation.}
\label{TAB_3}
\end{table}

\newpage

\begin{figure} 
\begin{center}
\setlength{\unitlength}{0.240900pt}
\ifx\plotpoint\undefined\newsavebox{\plotpoint}\fi
\sbox{\plotpoint}{\rule[-0.175pt]{0.350pt}{0.350pt}}%
\begin{picture}(1500,1619)(0,0)
\tenrm
\sbox{\plotpoint}{\rule[-0.175pt]{0.350pt}{0.350pt}}%
\put(264,158){\rule[-0.175pt]{282.335pt}{0.350pt}}
\put(264,158){\rule[-0.175pt]{4.818pt}{0.350pt}}
\put(242,158){\makebox(0,0)[r]{0}}
\put(1416,158){\rule[-0.175pt]{4.818pt}{0.350pt}}
\put(264,428){\rule[-0.175pt]{4.818pt}{0.350pt}}
\put(242,428){\makebox(0,0)[r]{0.2}}
\put(1416,428){\rule[-0.175pt]{4.818pt}{0.350pt}}
\put(264,697){\rule[-0.175pt]{4.818pt}{0.350pt}}
\put(242,697){\makebox(0,0)[r]{0.4}}
\put(1416,697){\rule[-0.175pt]{4.818pt}{0.350pt}}
\put(264,967){\rule[-0.175pt]{4.818pt}{0.350pt}}
\put(242,967){\makebox(0,0)[r]{0.6}}
\put(1416,967){\rule[-0.175pt]{4.818pt}{0.350pt}}
\put(264,1236){\rule[-0.175pt]{4.818pt}{0.350pt}}
\put(242,1236){\makebox(0,0)[r]{0.8}}
\put(1416,1236){\rule[-0.175pt]{4.818pt}{0.350pt}}
\put(264,1506){\rule[-0.175pt]{4.818pt}{0.350pt}}
\put(242,1506){\makebox(0,0)[r]{1}}
\put(1416,1506){\rule[-0.175pt]{4.818pt}{0.350pt}}
\put(264,158){\rule[-0.175pt]{0.350pt}{4.818pt}}
\put(264,113){\makebox(0,0){-10}}
\put(264,1486){\rule[-0.175pt]{0.350pt}{4.818pt}}
\put(444,158){\rule[-0.175pt]{0.350pt}{4.818pt}}
\put(444,113){\makebox(0,0){-8}}
\put(444,1486){\rule[-0.175pt]{0.350pt}{4.818pt}}
\put(625,158){\rule[-0.175pt]{0.350pt}{4.818pt}}
\put(625,113){\makebox(0,0){-6}}
\put(625,1486){\rule[-0.175pt]{0.350pt}{4.818pt}}
\put(805,158){\rule[-0.175pt]{0.350pt}{4.818pt}}
\put(805,113){\makebox(0,0){-4}}
\put(805,1486){\rule[-0.175pt]{0.350pt}{4.818pt}}
\put(985,158){\rule[-0.175pt]{0.350pt}{4.818pt}}
\put(985,113){\makebox(0,0){-2}}
\put(985,1486){\rule[-0.175pt]{0.350pt}{4.818pt}}
\put(1166,158){\rule[-0.175pt]{0.350pt}{4.818pt}}
\put(1166,113){\makebox(0,0){0}}
\put(1166,1486){\rule[-0.175pt]{0.350pt}{4.818pt}}
\put(1346,158){\rule[-0.175pt]{0.350pt}{4.818pt}}
\put(1346,113){\makebox(0,0){2}}
\put(1346,1486){\rule[-0.175pt]{0.350pt}{4.818pt}}
\put(264,158){\rule[-0.175pt]{282.335pt}{0.350pt}}
\put(1436,158){\rule[-0.175pt]{0.350pt}{324.733pt}}
\put(264,1506){\rule[-0.175pt]{282.335pt}{0.350pt}}
\put(1,832){\makebox(0,0)[l]{\shortstack{{\Large $Z_L$}}}}
\put(850,23){\makebox(0,0){{\Large $n_f$}}}
\put(264,158){\rule[-0.175pt]{0.350pt}{324.733pt}}
\put(1166,953){\raisebox{-1.2pt}{\makebox(0,0){$\Diamond$}}}
\put(985,859){\raisebox{-1.2pt}{\makebox(0,0){$\Diamond$}}}
\put(805,785){\raisebox{-1.2pt}{\makebox(0,0){$\Diamond$}}}
\put(625,751){\raisebox{-1.2pt}{\makebox(0,0){$\Diamond$}}}
\put(1166,940){\rule[-0.175pt]{0.350pt}{6.504pt}}
\put(1156,940){\rule[-0.175pt]{4.818pt}{0.350pt}}
\put(1156,967){\rule[-0.175pt]{4.818pt}{0.350pt}}
\put(985,841){\rule[-0.175pt]{0.350pt}{8.431pt}}
\put(975,841){\rule[-0.175pt]{4.818pt}{0.350pt}}
\put(975,876){\rule[-0.175pt]{4.818pt}{0.350pt}}
\put(805,772){\rule[-0.175pt]{0.350pt}{6.022pt}}
\put(795,772){\rule[-0.175pt]{4.818pt}{0.350pt}}
\put(795,797){\rule[-0.175pt]{4.818pt}{0.350pt}}
\put(625,744){\rule[-0.175pt]{0.350pt}{3.373pt}}
\put(615,744){\rule[-0.175pt]{4.818pt}{0.350pt}}
\put(615,758){\rule[-0.175pt]{4.818pt}{0.350pt}}
\end{picture}
\caption{The values of $Z_L$ as a function of the flavour number for $R_2=0.5$.}
\end{center}
\end{figure}

\newpage

\begin{figure} 
\begin{center}
\setlength{\unitlength}{0.240900pt}
\ifx\plotpoint\undefined\newsavebox{\plotpoint}\fi
\sbox{\plotpoint}{\rule[-0.175pt]{0.350pt}{0.350pt}}%
\begin{picture}(1500,1619)(0,0)
\tenrm
\sbox{\plotpoint}{\rule[-0.175pt]{0.350pt}{0.350pt}}%
\put(264,158){\rule[-0.175pt]{282.335pt}{0.350pt}}
\put(264,158){\rule[-0.175pt]{4.818pt}{0.350pt}}
\put(242,158){\makebox(0,0)[r]{0}}
\put(1416,158){\rule[-0.175pt]{4.818pt}{0.350pt}}
\put(264,428){\rule[-0.175pt]{4.818pt}{0.350pt}}
\put(242,428){\makebox(0,0)[r]{0.2}}
\put(1416,428){\rule[-0.175pt]{4.818pt}{0.350pt}}
\put(264,697){\rule[-0.175pt]{4.818pt}{0.350pt}}
\put(242,697){\makebox(0,0)[r]{0.4}}
\put(1416,697){\rule[-0.175pt]{4.818pt}{0.350pt}}
\put(264,967){\rule[-0.175pt]{4.818pt}{0.350pt}}
\put(242,967){\makebox(0,0)[r]{0.6}}
\put(1416,967){\rule[-0.175pt]{4.818pt}{0.350pt}}
\put(264,1236){\rule[-0.175pt]{4.818pt}{0.350pt}}
\put(242,1236){\makebox(0,0)[r]{0.8}}
\put(1416,1236){\rule[-0.175pt]{4.818pt}{0.350pt}}
\put(264,1506){\rule[-0.175pt]{4.818pt}{0.350pt}}
\put(242,1506){\makebox(0,0)[r]{1}}
\put(1416,1506){\rule[-0.175pt]{4.818pt}{0.350pt}}
\put(264,158){\rule[-0.175pt]{0.350pt}{4.818pt}}
\put(264,113){\makebox(0,0){-10}}
\put(264,1486){\rule[-0.175pt]{0.350pt}{4.818pt}}
\put(444,158){\rule[-0.175pt]{0.350pt}{4.818pt}}
\put(444,113){\makebox(0,0){-8}}
\put(444,1486){\rule[-0.175pt]{0.350pt}{4.818pt}}
\put(625,158){\rule[-0.175pt]{0.350pt}{4.818pt}}
\put(625,113){\makebox(0,0){-6}}
\put(625,1486){\rule[-0.175pt]{0.350pt}{4.818pt}}
\put(805,158){\rule[-0.175pt]{0.350pt}{4.818pt}}
\put(805,113){\makebox(0,0){-4}}
\put(805,1486){\rule[-0.175pt]{0.350pt}{4.818pt}}
\put(985,158){\rule[-0.175pt]{0.350pt}{4.818pt}}
\put(985,113){\makebox(0,0){-2}}
\put(985,1486){\rule[-0.175pt]{0.350pt}{4.818pt}}
\put(1166,158){\rule[-0.175pt]{0.350pt}{4.818pt}}
\put(1166,113){\makebox(0,0){0}}
\put(1166,1486){\rule[-0.175pt]{0.350pt}{4.818pt}}
\put(1346,158){\rule[-0.175pt]{0.350pt}{4.818pt}}
\put(1346,113){\makebox(0,0){2}}
\put(1346,1486){\rule[-0.175pt]{0.350pt}{4.818pt}}
\put(264,158){\rule[-0.175pt]{282.335pt}{0.350pt}}
\put(1436,158){\rule[-0.175pt]{0.350pt}{324.733pt}}
\put(264,1506){\rule[-0.175pt]{282.335pt}{0.350pt}}
\put(1,832){\makebox(0,0)[l]{\shortstack{{\Large $Z_L$}}}}
\put(850,23){\makebox(0,0){{\Large $n_f$}}}
\put(264,158){\rule[-0.175pt]{0.350pt}{324.733pt}}
\put(1166,1032){\raisebox{-1.2pt}{\makebox(0,0){$\Diamond$}}}
\put(985,930){\raisebox{-1.2pt}{\makebox(0,0){$\Diamond$}}}
\put(805,852){\raisebox{-1.2pt}{\makebox(0,0){$\Diamond$}}}
\put(625,786){\raisebox{-1.2pt}{\makebox(0,0){$\Diamond$}}}
\put(444,751){\raisebox{-1.2pt}{\makebox(0,0){$\Diamond$}}}
\put(1166,1024){\rule[-0.175pt]{0.350pt}{3.613pt}}
\put(1156,1024){\rule[-0.175pt]{4.818pt}{0.350pt}}
\put(1156,1039){\rule[-0.175pt]{4.818pt}{0.350pt}}
\put(985,921){\rule[-0.175pt]{0.350pt}{4.577pt}}
\put(975,921){\rule[-0.175pt]{4.818pt}{0.350pt}}
\put(975,940){\rule[-0.175pt]{4.818pt}{0.350pt}}
\put(805,839){\rule[-0.175pt]{0.350pt}{6.263pt}}
\put(795,839){\rule[-0.175pt]{4.818pt}{0.350pt}}
\put(795,865){\rule[-0.175pt]{4.818pt}{0.350pt}}
\put(625,775){\rule[-0.175pt]{0.350pt}{5.300pt}}
\put(615,775){\rule[-0.175pt]{4.818pt}{0.350pt}}
\put(615,797){\rule[-0.175pt]{4.818pt}{0.350pt}}
\put(444,741){\rule[-0.175pt]{0.350pt}{4.818pt}}
\put(434,741){\rule[-0.175pt]{4.818pt}{0.350pt}}
\put(434,761){\rule[-0.175pt]{4.818pt}{0.350pt}}
\end{picture}
\caption{The values of $Z_L$ as a function of the flavour number for $R_2=0.6$.}
\end{center}
\end{figure}

\end{document}